\documentclass[11pt,twoside,onecolumn,journal,compsoc]{IEEEtran}
\usepackage{amsmath,amssymb,latexsym}
\usepackage{pst-all}
\usepackage[varg]{pxfonts}
\usepackage{mathrsfs}

\newtheorem{theorem}{\bf Theorem}[section]

\newtheorem{example}[theorem]{\bf Example}

\newtheorem{algorithm}[theorem]{\bf Algorithm}

\renewcommand{\iff}{\Leftrightarrow}

\newcommand{\lra}{\leftrightarrow}
\renewcommand{\le}{\leqslant}
\renewcommand{\ge}{\geqslant}

\newcommand{\im}{\qopname\relax{no}{Im}}

\newcommand{\lBrack}{\lbrack\!\lbrack}
\newcommand{\rBrack}{\rbrack\!\rbrack}

\newcommand{\oobslash}{\circledbslash\kern-10.2pt\bigcirc}
\newcommand{\ooslash}{\oslash\kern-10.2pt\bigcirc}

\renewcommand{\IEEEQED}{\IEEEQEDclosed}
\parskip=\smallskipamount

\begin{document}

\title{\Large\bf Determinization of fuzzy automata by means of the degrees\\ of language inclusion%
\thanks{Research supported by Ministry Education and Science, Republic of Serbia, Grant No. 174013}}
\author{Ivana Mici\'c, Zorana Jan\v ci\'c, Jelena Ignjatovi\'c, and Miroslav \'Ciri\'c%
\thanks{I. Mici\'c, Z. Jan\v ci\'c, J. Ignjatovi\'c, and M. \'Ciri\'c are with the University of Ni\v s, Faculty of Sciences and Mathematics, Department of Computer Science, Vi\v segradska 33, 18000 Ni\v s, Serbia}
\thanks{E-mail: ivanajancic84@gmail.com, zoranajancic329@gmail.com, jelena.ignjatovic@pmf.edu.rs, miroslav.ciric@pmf.edu.rs}
\thanks{Manuscript received April, 2014}}

\date{}

\IEEEtitleabstractindextext{
\begin{abstract}
Determinization of fuzzy finite automata is understood here as a procedure of their conversion into equi\-valent crisp-deterministic fuzzy automata, which can  be  viewed as being deterministic with possibly infi\-nitely many states, but with fuzzy sets of terminal states. Particularly significant determinization methods are those  that provide a minimal crisp-deterministic fuzzy automaton equivalent to the original fuzzy finite automaton,  called canonization methods.~One canonization method for fuzzy finite automata,
the~Brzozowski type deter\-miniza\-tion, has been developed recently by Jan\v ci\'c and \'Ciri\'c in \cite{JC.14}. Here we provide another~canonization  method
for a fuzzy finite automaton $\cal A=(A,\sigma, \delta,\tau)$ over a complete residuated lattice
$\cal L$, based on the degrees of inclusion of the right fuzzy languages associated with states of $\cal A$ into the left derivatives of the fuzzy
language recognized by $\cal A$.~The proposed procedure terminates in a finite number of steps whenever the membership values~taken~by~$\delta $,~$\sigma $ and $\tau $ generate a finite subsemiring of the semiring reduct of $\cal L$.~This procedure is generally
faster than the Brzozowski type deter\-miniza\-tion, and if the basic operations in the residuated lattice $\cal L$ can be performed
in constant time, it has the same compu\-ta\-tional time as all other determinization procedures pro\-vided in \cite{ICB.08,JIC.11,JMIC.14}.
\end{abstract}

\begin{IEEEkeywords}
Fuzzy automaton; Complete residuated lattice; Crisp-deterministic automaton; Determinization; Inclusion degree;
\end{IEEEkeywords}}

\maketitle

\section{Introduction}

Many practical applications of automata require determinization, a procedure
of converting a nondetermi\-nistic finite automaton to an equivalent deterministic finite automaton. In the context of fuzzy automata, determinization is generally
understood as conversion of a fuzzy finite automaton to an equivalent crisp-deterministic fuzzy automaton, which can be viewed as being deterministic with possibly infinitely
many states, but with a fuzzy set of terminal states. The standard determinization method, known as the subset  construction, converts a nondeterministic automaton with $n$ states to an equivalent deterministic automa\-ton with up to $2^n$ states, and the fuzzy version of this construction may even give an infinite crisp-de\-ter\-ministic fuzzy automaton. That is why an extremely important task is to find such methods that will mitigate the potential enormous growth of the number of states during the determinization. Particularly signi\-ficant are those determinization methods that provide minimal crisp-de\-ter\-ministic fuzzy automata, called canoniza\-tion methods.
One canoniza\-tion method for fuzzy finite automata,
the Brzozowski type determi\-nization, has been developed recently by Jan\v ci\'c and \'Ciri\'c in \cite{JC.14}. The purpose of this paper is to pro\-vide another canonization method
for fuzzy finite automata over a complete residuated lattice.

Determinization of fuzzy automata was first studied by B\v elohl\'avek \cite{Bel.02b}, in the context of fuzzy finite automata over a complete distributive lattice, an Li and Pedrycz \cite{LP.05}, in the context of fuzzy finite automata over a lattice-ordered monoid. Determinization algorithms that were provided there generalize the subset construction. Another algorithm, provided by Ignjatovi\'c et al. \cite{ICB.08}, also generalizes the subset construction
and for any input produces a smaller crisp-deterministic fuzzy automaton than algorithms from \cite{Bel.02b,LP.05}. This crisp-deterministic fuzzy automaton can be alternatively constructed by means of the Nerode right con\-gruence of the original fuzzy finite automaton, and it was called in \cite{ICBP.10} the {\it Nerode automaton\/} of the original fuzzy finite auto\-maton. The Nerode automaton was constructed in \cite{ICB.08} for fuzzy finite automata over a complete re\-sid\-uated lattice, and it was noted that the identical construction can also be made in a more general context, for fuzzy finite automata over a lattice-ordered monoid, and even for weighted finite automata over a semi\-ring. The same construction was also transferred in \cite{CDIV.10} to weighted automata over strong bimonoids. The algorithm proposed by Jan\v ci\'c et al. in \cite{JIC.11}, which generalizes the ``transition sets construction``
given in \cite{vGP.08p,vGP.08tr}, produces a crisp-deterministic fuzzy or weighted automaton that is even smaller than the Nerode automaton, and further progress has been made in a recent paper by  Jan\v ci\'c et al. \cite{JMIC.14}, where algorithms which perform both determinization~and state re\-duc\-tion have been provided. In addition,  Jan\v ci\'c and \'Ciri\'c in \cite{JC.14} adapted the well-known Brzozowski's double reversal determinization algorithm to fuzzy auto\-mata and provided
a  Brzozowski type determini\-zation
algorithm that yields a minimal crisp-deterministic fuzzy automaton equivalent to the original fuzzy finite automaton.~It is important to note
that another form of determinization, more general than crisp-determinization,
has recently been studied by Gonz\'alez de Mendivil and Garitagoitia \cite{GdMG.14,GdMG.14b}.

In the present paper we propose another determinization method which for any fuzzy finite
automaton ${\cal A}=(A,\sigma,\delta,\tau )$ over a complete residuated lattice $\cal L$
produces a minimal crisp-deterministic fuzzy autom\-aton ${\cal A}_d$ equivalent to $\cal A$.~The automaton ${\cal A}_d$ does not necessarily have a finite number of states, but~when\-ever the membership values~taken~by~$\delta $,~$\sigma $ and $\tau $ generate a finite~sub\-semi\-ring of the semiring reduct of $\cal L$, then the number of states of ${\cal A}_d$ is also finite.~The proposed canonization procedure is based on the degrees of inclusion of the right fuzzy languages associated with states of $\cal A$ into the left derivatives of the fuzzy language recognized by $\cal A$. The computation time of this procedure is generally better than the computation time of the Brzozowski type determinization, and if the basic operations in the underlying residuated lattice can be performed
in constant time, it has the same computational time as all other determinization procedures pro\-vided in \cite{ICB.08,JIC.11,JMIC.14}.

The paper is organized as follows. In Section 2 we recall basic notions and notation concerning fuzzy sets and relations, fuzzy automata and languages, crisp-deterministic fuzzy automata, and the Nerode and reverse Nerode automaton.
Our main theoretical results are presented in Section 3, where for a given
fuzzy finite automaton we construct an equivalent accessible
crisp-deterministic fuzzy automaton whose states
are fuzzy sets defined by means of the degrees of inclusion of certain
fuzzy languages. In Section 4 we state an algorithm which constructs this
crisp-deterministic fuzzy automaton, perform the analysis of the computation
time, and give examples that demonstrates the application of the algorithm.

\section{Preliminaries}
\subsection{Fuzzy sets and relations}

In this paper we use complete residuated lattices as
structures of membership values. A {\it residuated lattice\/} is
an algebra ${\cal L}=(L,\land ,\lor , \otimes ,\to , 0, 1)$ such
that
\begin{itemize}
\parskip=0pt\itemindent=1.4mm
\item[{\rm (L1)}] $(L,\land ,\lor , 0, 1)$ is a lattice with the
least element $0$ and the greatest element $1$, \item[{\rm (L2)}]
$(L,\otimes ,1)$ is a commutative monoid with the unit $1$,
\item[{\rm (L3)}] $\otimes $ and $\to $ form an {\it adjoint
pair\/}, i.e., they satisfy the {\it adjunction property\/}: for
all $x,y,z\in L$,
\begin{equation}\label{eq:adj}
x\otimes y \le z \ \iff \ x\le y\to z .
\end{equation}
\end{itemize}
If, additionally, $(L,\land ,\lor , 0, 1)$ is a complete lattice, then ${\cal L}$ is called a {\it
complete residuated lattice\/}.

The algebra ${\cal L}^*=(L,\lor ,\otimes, 0, 1)$ is a semiring, and it is called the \emph{semiring reduct\/} of $\cal L$.

The operations $\otimes $ (called {\it multiplication\/}) and $\to
$ (called {\it residuum\/}) are intended for modeling the
conjunc\-tion and implication of the corresponding logical calculus,
and supremum ($\bigvee $) and infimum ($\bigwedge $) are intend\-ed
for modeling of the existential and general quantifier,
respectively. An operation $\lra $ defined by
\begin{equation}\label{eq:bires}
x\lra y = (x\to y) \land (y\to x),
\end{equation}
called {\it biresiduum\/} (or {\it biimplication\/}), is used for
modeling the equivalence of truth values. It can be easily
shown that with respect to $\le $, $\otimes $ is isotonic in
both arguments, $\to $ is isotonic in the second and antitonic in
the first argument, and for any $x,y,z\in L$  the following
hold:
\begin{eqnarray}
&& (x\to y)\otimes x \le y, \label{eq:FMP}\\
&& y\leqslant x\to (x\otimes y), \label{eq:yle}
\end{eqnarray}
For other properties of complete residuated lattices one can refer
to \cite{Bel.02,BV.05}.

The most studied and applied structures of truth values, defined
on the real unit interval $[0,1]$ with  $x\land y =\min
(x,y)$ and $x\lor y =\max (x,y)$, are the {\it {\L}ukasiewicz
structure\/} ($x\otimes y = \max(x+y-1,0)$, $x\to y=
\min(1-x+y,1)$), the {\it Goguen} ({\it product\/}) {\it
structure\/} ($x\otimes y = x\cdot y$, $x\to y= 1$ if $x\le y$
and $=y/x$ otherwise) and the {\it G\"odel structure\/} ($x\otimes
y = \min(x,y)$, $x\to y= 1$ if $x\le y$ and $=y$
otherwise).
Another important set of truth values is the set
$\{a_0,a_1,\ldots,a_n\}$, $0=a_0<\dots <a_n=1$, with $a_k\otimes
a_l=a_{\max(k+l-n,0)}$ and $a_k\to a_l=a_{\min(n-k+l,n)}$. A
special case of the latter algebras is the two-element Boolean
algebra of classical logic with the support $\{0,1\}$. The only
adjoint pair on the two-element Boolean algebra consists of the
classical conjunction and implication operations. This structure
of truth values is called the {\it Boolean structure\/}.

In the sequel $\cal L$ will be a complete residuated
lattice. A {\it fuzzy subset\/} of a set $A$ {\it over\/} ${\cal
L}$, or simply a {\it fuzzy subset\/} of $A$, is any mapping from
$A$ into $L$. Ordinary crisp subsets of $A$ are considered as
fuzzy subsets of $A$ taking membership values in the set
$\{0,1\}\subseteq L$. Let $f$ and $g$ be two fuzzy subsets of
$A$. The {\it equality\/} of $f$ and $g$ is defined as the usual
equality of mappings, i.e., $f=g$ if and only if $f(x)=g(x)$, for
every $x\in A$. The {\it inclusion\/} $f\le g$ is also defined
pointwise: $f\le g$ if and only if $f(x)\le g(x)$, for every $x\in
A$. Endowed with this partial order the set $L^A$ of all fuzzy
subsets of $A$ forms a complete residuated lattice, in which the
meet (intersection) $\bigwedge_{i\in I}f_i$ and the join (union)
$\bigvee_{i\in I}f_i$ of an arbitrary family $\{f_i\}_{i\in I}$ of
fuzzy subsets of $A$ are mappings from $A$ into $L$ defined by
\[
\left(\bigwedge_{i\in I}f_i\right)(x)=\bigwedge_{i\in I}f_i(x), \qquad \left(\bigvee_{i\in
I}f_i\right)(x)=\bigvee_{i\in I}f_i(x),
\]
and the {\it product\/} $f\otimes g$ is a fuzzy subset defined by
$f\otimes g (x)=f(x)\otimes g(x)$, for each $x\in A$. The \emph{degree of inclusion} $I(f,g)$ of two fuzzy sets $f,g\in L^A$, in that order (or the
degree of inclusion of $f$ into $g$),  is~defined~by
\[
I(f,g)= \bigwedge_{a\in A}f(a)\to g(a).
\]
In other words, $I(f,g)$ is a measure of ``how much  $f$ is contained in $g$''.

 A {\it fuzzy relation\/} between sets $A$ and $B$ (in this order) is any mapping from
$A\times B$ to $L$, i.e. , any fuzzy subset of $A\times B$, and the equality, inclusion (ordering), joins and meets
of fuzzy relations are defined as for fuzzy sets. Set of all fuzzy relations between $A$ and $B$ will be denoted by $L^{A\times B}$. In particular, a fuzzy relation on a set $A$ is any function from
$A\times A$ to $L$, i.e., any fuzzy subset of $A\times A$. The set of all fuzzy relations on $A$ will be denoted by $L^{A\times A}$. The \emph{reverse} or \emph{inverse} of a fuzzy relation $\alpha\in L^{A\times B}$ is a fuzzy relation $\alpha^{-1}\in L^{B\times A}$ defined by $\alpha^{-1}(b, a) = \alpha(a, b)$, for all $a\in A$ and $b\in B$. A crisp relation is a fuzzy
relation which takes values only in the set $\{0, 1\}$, and if $\alpha$ is a crisp relation of $A$ to $B$, then expressions
$"\alpha(a, b) = 1"$ and $"(a, b)\in \alpha"$ will have the same meaning.

For non-empty sets $A$, $B$ and $C$, and fuzzy relations $\alpha\in L^{A\times B}$ and $\beta\in L^{B\times C}$,  their {\it
composition\/} $\alpha\circ \beta\in L^{A\times C}$ is a fuzzy relation  defined by
\begin{equation}\label{eq:comp.rr}
(\alpha \circ \beta )(a,c)=\bigvee_{b\in B}\,\alpha(a,b)\otimes \beta(b,c),
\end{equation}
for all $a\in A$ and $c\in C$.
 For $f\in L^{A}$, $\alpha \in L^{A\times B}$ and $g\in L^{B}$, compositions $f\circ\alpha\in L^{B}$ and $\alpha\circ g\in L^{A}$ are fuzzy sets defined by
\begin{equation}\label{eq:comp.sr}
(f \circ \alpha)(b)=\bigvee_{a\in A}\,f(a)\otimes \alpha(a,b),\qquad
(\alpha \circ g)(a)=\bigvee_{b\in B}\,\alpha(a,b)\otimes g(b),
\end{equation}
for every $a\in A$ and $b\in B$.
Finally, the composition of two fuzzy sets $f,g\in L^{A}$ is an element $f\circ
g\in L$ (scalar) defined by
\begin{equation}\label{eq:comp.ss}
f \circ g =\bigvee_{a\in A}\,f(a)\otimes g(a) .
\end{equation}
When the underlying sets are finite, fuzzy relations can be interpreted as matrices and fuzzy sets as vectors with entries in $L$, and then the composition
of fuzzy relations can be interpreted as the matrix product, compositions
of fuzzy sets and fuzzy relations as vector-matrix products, and the composition
of two fuzzy set as the scalar (dot) product.

It is easy to verify that the composition of fuzzy relations is associative,
i.e.,
\begin{equation}
(\alpha\circ\beta)\circ\gamma = \alpha\circ(\beta\circ\gamma),  \label{eq:comp.as}
\end{equation}
for all $\alpha\in L^{A\times B}$, $\beta\in L^{B\times C}$ and $\gamma\in
L^{C\times D}$, and
\begin{equation}
( f \circ\alpha) \circ\beta= f\circ (\alpha\circ\beta),\quad ( f \circ\alpha) \circ g= f\circ (\alpha\circ g),\quad (\alpha\circ\beta)\circ h = \alpha\circ(\beta\circ h) \label{eq:comp.as2}
\end{equation}
for all $\alpha\in L^{A\times B}$, $\beta\in L^{B\times C}$, $f\in L^{A}$, $g\in L^{B}$ and $h\in L^{C}$. Hence, all parentheses in (\ref{eq:comp.as})
and (\ref{eq:comp.as2}) can be omitted.

\subsection{Fuzzy automata}

Throughout this paper, $\Bbb N$ denotes the set of natural numbers (without zero), $X$ is an (finite) alphabet, $X^+$ and $X^*$ denote, respectively, the free semigroup and the free monoid over $X$, $\varepsilon$ denotes the empty word in $X^*$, and if not noted otherwise, ${\cal L}$ is a complete residua\-ted lattice.

A {\it fuzzy automaton over\/} $\cal L$ and $X$, or simply a {\it fuzzy automaton\/}, is a quadruple
${\cal A}=(A,\sigma,\delta,\tau)$, where $A$ is a non-empty set, called  the
{\it set of states\/}, $\delta:A\times X\times A\to L$ is a
fuzzy subset of $A\times X\times A$, called the {\it fuzzy transition function\/}, and $\sigma: A\to L$ and $\tau : A\to L$ are fuzzy subsets of $A$, called the {\it fuzzy set of initial states} and the {\it fuzzy set terminal states}, respectively. We can
interpret $\delta (a,x,b)$ as the degree to which an input letter $x\in X$
causes a transition from a state $a\in A$ into a
state $b\in A$, and we can interpret $\sigma(a)$ and $\tau(a)$ as the degrees  to which $a$ is respectively an input state and a terminal state. For methodological reasons we allow the set of states $A$ to be infinite. A fuzzy auto\-maton whose set of states is finite is called a {\it fuzzy finite automaton\/}. A fuzzy automaton over the Boolean structure is called a \emph{nondeterministic automaton\/} or a \emph{Boolean automaton\/}.

Define a family $\{\delta_x\}_{x\in X}$ of fuzzy relations on $A$ by $\delta_x(a,b) = \delta(a,x,b)$, for each $x\in X$, and all $a,b\in A$, and extend this
family to the family $\{\delta_u\}_{u\in X^*}$ inductively, as follows:
$\delta_\varepsilon=\Delta_A$, where $\Delta_A$ is the crisp equality relation
on $A$, and
\begin{equation}\label{eq:du}
 \delta_{x_1x_2\dots x_n}=\delta_{x_1}\circ
\delta_{x_2}\circ\dots\circ \delta_{x_n}
\end{equation}
for all $n\in \Bbb N$, $x_1,x_2,\ldots ,x_n\in X$. Members of this family
are called {\it fuzzy transiton relations\/} of $\cal A$. Evidently, $\delta_{uv}= \delta_u\circ \delta_v$, for all $u,v\in X^*$. In addition, define
families $\{\sigma_u\}_{u\in X^*}$ and $\{\tau_u\}_{u\in X^*}$ by
\begin{equation}\label{eq:su.tu}
\sigma_u=\sigma\circ \delta_u, \qquad \tau_u=\delta_u\circ \tau,
\end{equation}
for all $u\in X^*$.
In dealing with fuzzy finite automata, fuzzy transition relations $\{\delta_u\}_{u\in X^*}$ are represented by fuzzy matrices with entries in $L$, whereas fuzzy
sets  $\{\sigma_u\}_{u\in X^*}$ are represented by row vectors  and $\{\tau_u\}_{u\in X^*}$ by column vectors with entries in $L$.

A {\it fuzzy language\/} in $X^*$ over ${\cal L}$, or
just a {\it fuzzy language\/}, is any fuzzy subset of $X^*$, i.e., any function from
$X^*$ into $L$. A {\it fuzzy language recognized by a fuzzy automaton\/} ${\cal A}=(A,\sigma,\delta , \tau )$ is a fuzzy language $\lBrack{\cal A}\rBrack \in L^{X^*}$ defined by
\begin{equation}\label{eq:recog}
\lBrack{\cal A}\rBrack(u) = \bigvee_{a,b\in A} \sigma (a)\otimes \delta_u(a,b)\otimes \tau(b)=\sigma \circ \delta_u\circ \tau  ,
\end{equation}
for any $u\in X^*$. In other words, the membership degree of the word
$u$ to the fuzzy language $\lBrack{\cal A}\rBrack$ is equal to the degree to which $\cal A$ recognizes or accepts
the word $u$. Fuzzy automata $\cal A$ and $\cal B$ are called   {\it language equivalent\/}, or just {\it equivalent\/}, if $\lBrack{\cal A}\rBrack=\lBrack{\cal B}\rBrack$.

The \emph{right
fuzzy language} associated with a state $a\in A$ is a fuzzy language $\tau_{a}\in L^{X^*}$ defined by
\[
\tau_{a}(u)=\tau_{u}(a)=\bigvee_{b\in A} \delta_u(a,b)\otimes\tau(b),
\]
for all $u\in X^*$, i.e., $\tau_{a}$ is the fuzzy language recognized by
a fuzzy automaton ${\cal A}_a=(A,\delta,a,\tau)$ obtained from ${\cal A}$ by replacing $\sigma $ with the single crisp initial state $a$.

Cardinality of a fuzzy automaton ${\cal
A}=(A,\sigma,\delta,\tau)$, in notation $|{\cal A} |$, is defined as the cardinality of its
set of states $A$. A fuzzy automaton ${\cal A}$ is called {\it minimal fuzzy automaton} of a fuzzy language $f \in L^{X^*}$ if $\lBrack {\cal A}\rBrack=f $
and $|{\cal A} |\leqslant|{\cal A'}|$, for every fuzzy automaton ${\cal A'}$ such that
$\lBrack {\cal A}'\rBrack=f $. A minimal fuzzy automaton recognizing a given
fuzzy language $f$ is not necessarily unique up to an isomorphism. This is also true for
nondeterministic automata.

Let ${\cal A}=(A,\delta,\sigma,\tau)$ be a fuzzy automaton over $\cal L$ and $X$. The {\it reverse fuzzy
automaton\/} of ${\cal A}$ is a fuzzy auto\-maton $\overline{\cal
A}=(A,\bar{\delta},\bar{\sigma},\bar{\tau})$, where $\bar{\sigma}=\tau$, $\bar{\tau}=\sigma$, and $\bar{\delta}:
A\times X\times A \to L$ is defined by $\bar{\delta}^A(a,x,b)=\delta^A(b,x,a)$,
for all $a,b\in A$ and $x\in X$. Roughly speaking, the reverse fuzzy automaton $\overline{\cal A}$ is obtained from ${\cal A}$ by exchanging  fuzzy sets of initial and final states and ``reversing'' all the transitions. Due to
the fact that the multipli\-cation $\otimes$ is commutative, we have that $\bar{\delta}_{u}(a,b)=\delta_{\bar{u}}(b,a)$, for all $a,b\in A$ and $u\in X^*$.

The {\it reverse fuzzy language\/} of a fuzzy language $f\in L^{X^*}$ is a fuzzy language $\overline{f}\in L^{X^*}$ defined by $\overline{f}(u)=f(\bar{u})$, for each $u\in X^*$. As $\overline{(\bar{u})} = u$ for all $u\in X^*$, we have that $\overline{(\overline{f})}=f$, for any fuzzy language $f$. It is easy to see that the reverse fuzzy automaton $\overline{\cal A}$ recognizes the reverse fuzzy language $\overline{\lBrack{\cal A}\rBrack}$ of the fuzzy language ${\lBrack{\cal A}\rBrack}$ recognized by $\cal A$, i.e., $\lBrack\overline{{\cal A}}\rBrack=\overline{\lBrack{\cal A}\rBrack}$.

We can visualize a fuzzy finite automaton ${\cal A}=(A,\sigma,\delta,\tau )$ representing it as a labelled directed graph whose nodes are states of $\cal A$, an edge from a node $a$ to a node $b$ is labelled by pairs of the form $x/\delta_x(a,b)$, for any $x\in X$, and for any node $a$ we draw an arrow labelled by $\sigma(a)$ that enters this node, and an arrow labelled by $\tau(a)$ coming out of this node. For the sake of simplicity, we do not draw edges whose all labels are of the form $x/0$, and incoming and outgoing arrows labelled by $0$. In particular, if $\cal A$ is a Boolean auto\-maton, instead of any label of the form $x/1$ we write just $x$, initial states are marked by incoming arrows without any label, and terminal states are marked by double circles.

For more information on fuzzy automata over complete residuated lattices we refer to \cite{IC.10,ICB.08,ICBP.10,Qiu.01,Qiu.02,Qiu.06,WQ.10,XQL.09}.

\subsection{Crisp-deterministic fuzzy automata}\label{sec:cdfa}

Let ${\cal A}=(A,\sigma,\delta,\tau)$ be a fuzzy automaton over  $X$ and  ${\cal L}$. The fuzzy transition function $\delta $ is called {\it crisp-deter\-ministic\/} if for every $x\in X$ and every $a\in A$ there exists
$a'\in A$ such that $\delta_x(a,a')=1$, and $\delta_x(a,b)=0$, for all $b\in A\setminus
\{a'\}$. The fuzzy set of initial states $\sigma $ is called {\it crisp-deterministic\/}
if there exists $a_0\in A$ such that $\sigma(a_0)=1$, and $\sigma(a)=0$, for every
$a\in A\setminus \{a_0\}$. If both $\sigma $ and $\delta $ are crisp-deterministic,
then $\cal A$ is called a {\it crisp-deterministic fuzzy automaton\/} (for short: {\it
cdfa\/}), and if it is finite, then it is called a {\it crisp-deterministic fuzzy finite automaton\/} (for short: {\it
cdffa\/}).

A crisp-deterministic fuzzy automaton can also be defined as a quadruple ${\cal A}=(A,\delta,a_0,\tau )$, where $A$ is a non-empty {\it set of states\/}, $\delta :A\times X\to A$ is a {\it transition function\/},
$a_0\in A$ is an {\it initial state\/} and $\tau\in L^A$ is a {\it fuzzy set of terminal states\/}. The
transition function $\delta $ can be extended to a function $\delta^*:A\times X^*\to A$ in the following way:
$\delta^*(a,\varepsilon)=a$, for every $a\in A$, and $\delta^*(a,ux)=\delta(\delta^*(a,u),x)$, for
all $a\in A$, $u\in X^*$ and $x\in X$. In this case the fuzzy language $\lBrack {\cal A}\rBrack$ recognized by  $\cal
A$ is given~by
\begin{equation}\label{eq:cd-beh}
\lBrack {\cal A}\rBrack(u)=\tau (\delta^*(a_0,u)),
\end{equation}
for every $u\in X^*$. Clearly, the image of $\lBrack {\cal A}\rBrack$ is contained in the image
of $\tau $ which is finite if the set of states $A$ is finite. A fuzzy language $f : X^*\to L$ is called {\it cdffa-recognizable\/} if
there is a crisp-deterministic fuzzy finite auto\-maton $\cal A$ over $X$ and ${\cal L}$ such that
$\lBrack {\cal A}\rBrack=f $. Then we say that $\cal A$ {\it recognizes\/} $f $.

A state $a\in A$ is called {\it accessible\/} if there exists $u\in X^*$ such that
$\delta^*(a_0,u)=a$. If every state of $\cal A$ is acces\-sible, then $\cal A$ is called an {\it
accessible crisp-deterministic fuzzy automaton\/}.

The initial state and transitions of a crisp-deterministic fuzzy automaton are graphically represented as in the case of Boolean automata, and the fuzzy set of terminal states is represented as in the case of fuzzy finite automata.

For a fuzzy language $f\in L^{X^*}$ and $u\in X^*$, we define a fuzzy language $f_u\in L^{X^*}$ by $f_u(v)=f(uv)$, for every $v\in X^*$. The fuzzy language $f_u $ is commonly called the {\it left derivative\/} of $f $ with respect to $u$, but for the sake of simplicity, $f_u$ will be called simply the \emph{derivative}
of $f $ with respect to $u$. Let $A_f =\{f_u\mid u\in X^*\}$
denote the set of all derivatives of $f $, and let $\delta_f :A_f \times X\to
A_f $ and $\tau_f :A_f\to L$  be functions defined by
\begin{equation}\label{eq:delta-tau.phi}
\delta_f(g ,x)=g_{x} \ \ \ \text{and}\ \ \ \tau_f(g )=g (\varepsilon),
\end{equation}
for all $g \in A_f $ and $x\in X$. Then ${\cal A}_f = (A_f , \delta_f
,f ,\tau_f )$ is an accessible  crisp-deterministic fuzzy automaton, and
it is called the  {\it derivative automaton\/} of the fuzzy language $f $ \cite{IC.10,ICBP.10}. It was proved in
\cite{ICBP.10} that the derivative automaton ${\cal A}_f $ is a minimal  crisp-deterministic fuzzy auto\-maton which recognizes $f$, and therefore, ${\cal A}_f$ is finite if and only if the fuzzy language $f $ is cdffa-recognizable. An algorithm for construction of the
derivative automaton of a fuzzy language, based on simultaneous construction of the derivative automata of
ordinary languages $f^{-1}(a)$, for all $a\in \im (f )$,  was also given in \cite{ICBP.10}.

A crisp-deterministic fuzzy automaton ${\cal A}$ is called a {\it minimal crisp-deterministic fuzzy automaton} of a fuzzy language $f$ if $\lBrack {\cal A}\rBrack=f$ and $|{\cal A} |<|{\cal A'}|$, for any crisp-deterministic fuzzy automaton ${\cal A'}$ such that $\lBrack {\cal A}'\rBrack=f$.

Further, the \textit{Nerode automaton} of a fuzzy automaton ${\cal A}=(A,\sigma,\delta,\tau)$ is a crisp-deterministic fuzzy autom\-aton ${\cal A}_{N}=(A_{N},\sigma_{\varepsilon},\delta_{N},\tau_{N})$,
where $A_{N}=\{\sigma_{u}\mid u\in X^{*}\}$, and $\delta_{N}:A_{N}\times X\longrightarrow A_{N}$ and $\tau_{N}:A_{N}\to L$ are defined, for all $u\in X^*$ and $x\in X$, by
\[
\delta_{N}(\sigma_{u},x)=\sigma_{ux},\qquad\qquad \tau_{N}(\sigma_{u})=\sigma_{u}\circ\tau,
\]
 The Nerode automaton was first constructed in \cite{ICB.08}, where it was shown that the Nerode automaton of  $\cal A$ is equivalent to $\cal A$, i.e., $\lBrack{\cal A}_{N}\rBrack={\lBrack{\cal A}\rBrack}$. The name "Nerode autom\-aton" was introdu\-ced in \cite{ICBP.10}.

The \textit{reverse Nerode automaton} of a ${\cal A}$ is the Nerode automaton
of the reverse automaton of $\cal A$, i.e., a crisp-deterministic fuzzy automaton ${\cal A}_{\overline N}=(A_{\overline N},\tau_\varepsilon,\delta_{\overline N},\tau_{\overline N})$, where $A_{\overline N}=\{\tau_u\mid u\in X^*\}$, and $\delta_{\overline N}:A_{\overline N}\times X\to A_{\overline N}$ and $\tau_{\overline N}:A_{\overline N}\times L$ are defined by
\[
\delta_{\overline N}(\tau_u,x)=\tau_{xu},\qquad\qquad \tau_{\overline N}(\tau_u)=\sigma\circ \tau_u,
\]
for all $u\in X^*$ and $x\in X$. As shown in \cite{JC.14}, the reverse Nerode automaton of $\cal A$ is equivalent to the reverse fuzzy automaton of $\cal A$, i.e., $\lBrack{\cal A}_{\overline N}\rBrack=\overline{\lBrack{\cal A}\rBrack}$.

\section{The main results}

Let ${\cal A}=(A,\delta,\sigma,\tau)$ be a fuzzy automaton over an alphabet
$X$ and a complete residuated lattice ${\cal L}$.~We~define inductively a family $\{d_u\}_{u\in X^*}$ of fuzzy subsets of $A$ as follows: for the
empty word $\varepsilon$ and $a\in A$~we~set
\begin{equation}\label{eq:de.varepsilon}
d_{\varepsilon}(a)=\bigwedge_{w\in X^*}\tau_{w}(a)\to\sigma\circ\tau_{w},
\end{equation}
and for all $u\in X^*$, $x\in X$ and $a\in A$ we set
\begin{equation}\label{eq:de.ux}
d_{ux}(a)=\bigwedge_{w\in X^*}\tau_{w}(a)\to d_{u}\circ\delta_{x}\circ\tau_{w}=\bigwedge_{w\in X^*}\tau_{w}(a)\to d_{u}\circ\tau_{xw}.
\end{equation}

The following theorem is one of the most important result of this paper.

\begin{theorem}\label{th:dsigma}\it
Let ${\cal A}=(A,\delta,\sigma,\tau)$ be a fuzzy automaton over an alphabet
$X$ and a complete residuated lattice ${\cal L}$. Then for all $u,v\in X^*$ we have
\begin{equation}\label{eq:de.tau.sigma.tau}
d_{u}\circ\tau_{v}=\sigma_{u}\circ\tau_{v}.
\end{equation}

\end{theorem}

\begin{IEEEproof}
By induction on the length of the word  $u$ we will prove that (\ref{eq:de.tau.sigma.tau})
is true for every $v\in X^*$. First, for the empty word $\varepsilon $ we have that
\[
d_{\varepsilon}(a)=\bigwedge_{w\in X^{*}}\tau_{w}(a)\rightarrow\sigma\circ\tau_{w}\le\tau_{v}(a)\rightarrow\sigma\circ\tau_{v}
\]
holds for all $a\in A$ and $v\in X^*$, and according to (\ref{eq:FMP}),
\[
d_{\varepsilon}\circ\tau_{v}=\bigvee_{a\in A}d_{\varepsilon}(a)\otimes\tau_{v}(a)\le\bigvee_{a\in A}(\tau_{v}(a)\rightarrow\sigma\circ\tau_{v})\otimes\tau_{v}(a)\le\sigma\circ\tau_{v}.
\]
Therefore, $d_{\varepsilon}\circ\tau_{v}\le\sigma\circ\tau_{v}$,
for all $v\in X^*$.

Also, according to (\ref{eq:yle}) and the fact that $\rightarrow$ is isotone in the second argument we have that
\[
\begin{aligned}
d_{\varepsilon}(a)&=\bigwedge_{w\in X^{*}}\tau_{w}(a)\rightarrow\sigma\circ\tau_{w} = \bigwedge_{w\in X^{*}}\bigl(\tau_{w}(a)\rightarrow\bigl(\bigvee_{b\in X^*}\sigma(b)\otimes\tau_{w}(b)\bigr)\bigr)\\
&\ge\bigwedge_{w\in X^{*}}\tau_{w}(a)\rightarrow(\sigma(a)\otimes\tau_{w}(a))\ge\sigma(a),
\end{aligned}
\]
 for all $a\in A$, so
\begin{equation*}\label{eq:de2}
d_{\varepsilon}\circ\tau_{v}=\bigvee_{a\in A}d_{\varepsilon}(a)\otimes\tau_{v}(a)\ge\bigvee_{a\in A}\sigma(a)\otimes\tau_{v}(a)=\sigma\circ\tau_{v}.
\end{equation*}
Therefore, $d_{\varepsilon}\circ\tau_{v}=\sigma\circ\tau_{v}$, for all $v\in X^*$, which means that (\ref{eq:de.tau.sigma.tau})
holds for $u=\varepsilon $ and every $v\in X^*$.

Let $u\in X^*$ be a word such that (\ref{eq:de.tau.sigma.tau}) holds for every word $v\in X^*$, and consider an arbitrary $x\in X$. Then
\[
d_{ux}(a)=\bigwedge_{w\in X^{*}}\tau_{w}(a)\rightarrow d_{u}\circ\delta_{x}\circ\tau_{w}\le\tau_{v}(a)\rightarrow d_{u}\circ\delta_{x}\circ\tau_{v},
\]
 for all $a\in A$ and $v\in X^*$, and according to (\ref{eq:FMP}),
\[
\begin{aligned}
d_{ux}\circ\tau_{v}&=\bigvee_{a\in A}d_{ux}(a)\otimes\tau_{v}(a)\le\bigvee_{a\in A}\bigl(\tau_{v}(a)\rightarrow d_{u}\circ\delta_{x}\circ\tau_{v}\bigr)\otimes\tau_{v}(a)\\
&\le d_{u}\circ\delta_{x}\circ\tau_{v}=d_{u}\circ\tau_{xv}=\sigma_{u}\circ\tau_{xv}=\sigma_{ux}\circ\tau_{v}.
\end{aligned}
\]
Next, according to (\ref{eq:yle}) and the fact that $\rightarrow$ is isotone in the second argument we have that
\[
\begin{aligned}
d_{ux}(a)&=\bigwedge_{w\in X^{*}}\tau_{w}(a)\rightarrow d_{u}\circ\delta_{x}\circ\tau_{w} = \bigwedge_{w\in X^{*}}\bigl(\tau_{w}(a)\rightarrow\bigl(\bigvee_{b\in X^*}(d_{u}\circ\delta_{x})(b)\otimes\tau_{w}(b)\bigr)\bigr)\\
&\ge\bigwedge_{w\in X^{*}}\tau_{w}(a)\rightarrow\bigl((d_{u}\circ\delta_{x})(a)\otimes\tau_{w}(a)\bigr)\ge(d_{u}\circ\delta_{x})(a),
\end{aligned}
\]
for all $a\in A$, so
\[\begin{aligned}\label{eq:du2}
d_{ux}\circ\tau_{v}&=\bigvee_{a\in A}d_{ux}(a)\otimes\tau_{v}(a)\ge\bigvee_{a\in A}(d_{u}\circ\delta_{x})(a)\otimes\tau_{v}(a)\\ &=d_u\circ\delta_{x}\circ\tau_{v}=d_{u}\circ\tau_{xv}=\sigma_{u}\circ\tau _{xv}=\sigma_{ux}\circ\tau _{v},
\end{aligned}
\]
and hence, $d_{ux}\circ\tau_{v}=\sigma_{ux}\circ\tau _{v}$. Thus, we conclude
that (\ref{eq:de.tau.sigma.tau}) holds for all $u,v\in X^*$.
\end{IEEEproof}

\smallskip

According to the previous theorem we have that
\[
d_{\varepsilon}(a)=\bigwedge_{w\in X^{*}}\tau_{w}(a)\rightarrow\sigma\circ\tau_{w}
\]
and
\[
d_{ux}(a)=\bigwedge_{w\in X^{*}}\tau_{w}(a)\rightarrow
d_{u}\circ\tau_{xw}=\bigwedge_{w\in X^{*}}\tau_{w}(a)\rightarrow
\sigma_{u}\circ\tau_{xw}=\bigwedge_{w\in X^{*}}\tau_{w}(a)\rightarrow \sigma_{ux}\circ\tau_{w}
\]
for all $a\in A$, $u\in X^*$ and $x\in X$, and hence
\begin{equation}\label{eq:for.min}
d_{u}(a)=\bigwedge_{w\in X^{*}}\tau_{w}(a)\rightarrow\sigma_{u}\circ\tau_{w}=\bigwedge_{w\in X^{*}}\tau_{a}(w)\rightarrow f_u(w)=I(\tau_a,f_u)
\end{equation}
holds for all $u\in X^*$ and $a\in A$, where $f=\lBrack{\cal A}\rBrack $.~In
other words, for all $u\in X^*$ and $a\in A$ we can~understand $d_u(a)$ as the degree of inclusion of the right
fuzzy language $\tau_a$ into the left derivative $f_u$ of the fuzzy language
$f$ recognized by $\cal A$.

Now,  set $A_{d}=\{d_{u}\mid u\in X^{*}\}$ and define functions $\delta_{d}:A_{d}\times X\rightarrow A_{d}$
and $\tau_{d}:A_{d}\rightarrow L$ by
\[
\delta_{d}(d_{u},x)=d_{ux},\qquad \qquad \tau_{d}(d_{u})=d_{u}\circ\tau,
\]
for all $d_{u}\in A_{d}$ and $x\in X$.

We have the following:

\begin{theorem}\label{th:cale}\it
Let ${\cal A}=(A,\delta,\sigma,\tau)$ be a fuzzy automaton over an alphabet $X$ and a complete residuated lattice ${\cal L}$. Then    ${\cal A}_{d}=(A_{d},\delta_{d},d_{\varepsilon},\tau_{d})$ is an accessible crisp-deter\-ministic fuzzy automaton equivalent to~${\cal A}$.
\end{theorem}

\begin{IEEEproof}
Let $u,v\in X^*$ such that $d_u=d_v$. Then for every $x\in X$ and $a\in A$ we have that
\[
d_{ux}(a)=\bigwedge_{w\in X^{*}}\tau_{w}(a)\rightarrow d_{u}\circ\delta_{x}\circ\tau_{w}=\bigwedge_{w\in X^{*}}\tau_{w}(a)\rightarrow d_{v}\circ\delta_{x}\circ\tau_{w}=d_{vx}(a),
\]
so $d_{ux}=d_{vx}$, and hence, $\delta_{d} $ is a well-defined mapping. It is evident that $\tau_{d}$ is also
a well-defined mapping. Thus, we have that
${\cal A}_{d}=(A_{d},\delta_{d},d_e,\tau_{d})$ is an accessible crisp-deterministic fuzzy automaton. According to Theorem \ref{th:dsigma} and definitions of fuzzy languages recognized by a fuzzy automaton and a crisp-determi\-nistic fuzzy automaton we have that
\[
\lBrack {\cal A}_d\rBrack(u)=\tau_d \bigl(\delta_d^*(d_\varepsilon,u)\bigr) = \tau_d (d_u) = d_u\circ\tau =
\sigma_u\circ \tau = \lBrack {\cal A}\rBrack (u),
\]
for every $u\in X^*$, and we have proved that ${\cal A}_d$ is equivalent to $\cal A$.
\end{IEEEproof}

The next theorem establishes the minimality of ${\cal A}_{d}$.

\begin{theorem}\label{th:cam}\it
Let ${\cal A}=(A,\delta,\sigma,\tau)$ be a fuzzy automaton over an alphabet $X$ and a complete residuated lattice ${\cal L}$. Then, ${\cal A}_{d}$ is a minimal crisp-deterministic fuzzy au\-tom\-aton equivalent to ${\cal A}$.
\end{theorem}

\begin{IEEEproof}
For the sake of simplicity set $\lBrack {\cal A}\rBrack = f$. According to Theorem 4.1 \cite{ICBP.10}, the derivative automaton ${\cal A}_{f}$ of $f$ is a minimal crisp-deterministic fuzzy\ automaton recognizing $f$, i.e., a minimal crisp-deterministic fuzzy
automaton equivalent to ${\cal A}$. Therefore, in order to show that the
automaton ${\cal A}_{d}=(A_{d},\delta_{d},d_{\varepsilon},\tau_{d})$ is a
minimal crisp-deterministic fuzzy automaton equivalent to ${\cal A}$ it is enough to prove that there exists a surjective mapping from ${\cal A}_{f}$ to ${\cal A}_{d}$.

Let $\phi : A_{f}\to A_{d}$ be a mapping defined by $\phi(f _u)= d_{u}$. By (\ref{eq:for.min}), for any $u,v \in X^{*}$ such that $f_u=f_v$, we have
\[
\sigma_{u}\circ\tau_{w}=\lBrack {\cal A}\rBrack(uw)=\lBrack {\cal A}\rBrack_u(w)=f_u(w)=f_v(w)=\lBrack {\cal A}\rBrack_v(w)=\lBrack {\cal A}\rBrack(vw)=\sigma_{v}\circ\tau_{w},
\]
for every $w\in X^{*}$. Consequently, for any $a\in A$ we have that
\[
d_{u}(a)=\bigwedge_{w\in X^{*}}\tau_{w}(a)\rightarrow\sigma_{u}\circ\tau_{w}=\bigwedge_{w\in X^{*}}\tau_{w}(a)\rightarrow\sigma_{v}\circ\tau_{w}=d_{v}(a),
\]
and hence, $d_u=d_v$. This means that $\phi$ is a well defined mapping. It is clear that $\phi$ is surjective. Thus, ${\cal A}_{d}$ is a minimal crisp-deterministic fuzzy automaton equivalent to ${\cal A}$.
\end{IEEEproof}

Let $\psi $ be a fuzzy relation on the set of states of a fuzzy automaton ${\cal A}=(A,\sigma,\delta,\tau)$.~According to definitions provided in \cite{JMIC.14}, $A^\psi=\{\psi^u\,|\,u\in X^*\}$ is a collection of fuzzy subsets of $A$
given by
\begin{equation}\label{eq:psi.u}
\psi^\varepsilon = \psi \circ \tau, \quad \psi^{xu}=\psi\circ \delta_x\circ
\psi^u, \ \ \text{for all $u\in X^*$, $x\in X$,}
\end{equation}
and ${\cal A}^\psi = (A^\psi ,\psi^\varepsilon,\delta^\psi ,\tau^\psi)$ is
a crisp-deterministic fuzzy automaton whose transition function~and fuzzy set of terminal states are given by
\begin{equation}\label{eq:a.psi}
\delta^\psi (\psi^u,x)=\psi^{xu},\quad \tau^\psi (\psi^u)=\sigma\circ \psi^u,
\ \ \text{for all $u\in X^*$, $x\in X$.}
\end{equation}
The fuzzy relation $\psi $ is called {\it left invariant\/} if $\sigma\circ\psi \leqslant \sigma$ and $\delta_x\circ\psi \leqslant \psi\circ \delta_x$,~for each $x\in X$, and it is called {\it weakly left invariant\/} if $\sigma_u\circ\psi \leqslant \sigma_u$, for each $u\in X^*$.~Clearly, every left invariant fuzzy relation is weakly left invariant.~If $\psi $ is reflexive and weakly left invariant, then ${\cal A}^\psi $ is equivalent to $\cal A$, and besides, $\sigma_u\circ\psi = \sigma_u$ and $\sigma_u\circ \psi^v=\sigma_u\circ
\tau_v$, for all $u,v\in X^*$.~Moreover, if $\psi^{\textrm{wli}} $ is the greatest weakly
left invariant fuzzy relation and $\psi^{\textrm{li}}$ is the greatest
left invariant fuzzy relation~on $\cal A$, then ${\cal A}^{\psi^{\textrm{wli}}} \leqslant {\cal A}^{\psi^{\textrm{li}}} \leqslant {\cal A}$ (cf.~\cite{JMIC.14}).

Now, define inductively a family $\{\Delta_u\}_{u\in X^*}$~of~fuzzy~subsets of $A$, as follows:
\begin{equation}\label{eq:De.varepsilon}
\Delta_\varepsilon(a)=\bigwedge_{w\in X^*}\psi^{w}(a)\to\sigma\circ\psi^{w},
\end{equation}
for all $a\in A$, and for all $u\in X^*$ and $x\in X$ we set
\begin{equation}\label{eq:De.ux}
\Delta_{ux}(a)=\bigwedge_{w\in X^*}\psi^{w}(a)\to \Delta_{u}\circ\psi^{xw}=\bigwedge_{w\in X^*}\psi^{w}(a)\to \Delta_{u}\circ\psi\circ
\delta_{x}\circ\psi^{w},
\end{equation}
for all $a\in A$. It is easy to check that Theorems \ref{th:dsigma}, \ref{th:cale} and \ref{th:cam} remain valid when $d_u$ is replaced by $\Delta_u$, which can significantly improve our canonization method since the cardinality of the family $\{\psi^w\}_{w\in X^*}$ is smaller than or equal to the cardinality of $\{\tau_w\}_{w\in X^*}$, and it may be~significantly smaller.~Furthermore, even in some cases where the family $\{\tau_w\}_{w\in X^*}$ is infinite, the family $\{\psi^w\}_{w\in X^*}$ may be finite (see Example 4.13 \cite{JMIC.14}).

\section{Algorithm and computational examples}\label{sec:6.3}

In order to compute the members of the collection of fuzzy sets $\{d_{u}\}_{u\in X^*}$, using formulas (\ref{eq:de.varepsilon}) and (\ref{eq:de.ux}), we first need to compute all members of the family  $\{\tau_{w}\}_{w\in X^*}$, what is nothing but the construction of the re\-verse Nerode automaton ${\cal A}_{\overline N}$ of the fuzzy automaton $\cal A$ which we want to determinize. Therefore, we can take the construction of  ${\cal A}_{\overline N}$ as the  first
step in the construction of  ${\cal A}_d$, and then we can proceed with the
computation of the collection $\{d_{u}\}_{u\in X^*}$.

The automaton ${\cal A}_{\overline N}$ can be computed using an algorithm which is derived from Algorithm 4.2 \cite{JMIC.14} as its particular case. For the sake of completeness we explicitly formulate this algorithm.

\begin{algorithm}[{{\emph{Construction of the reverse Nerode automaton ${\cal A}_{ \overline N}$}}}]\label{alg:A.reversNerode}\rm The input of this algorithm is a fuzzy finite autom\-aton ${\cal A}=(A,\delta,\sigma,\tau )$ with $n$ states, over a finite alphabet $X$ with $m$ letters and a complete
residuated lattice $\cal L$, and the output is the crisp-deterministic automaton ${\cal A}_{\overline N}=(A_{\overline N},\delta_{\overline N},\tau_\varepsilon,\tau_{\overline N})$.

The procedure is to construct the \emph{transition tree} of ${\cal A}_{\overline N}$ directly from ${\cal A}$, and during this procedure we use pointers $s(\cdot)$ which points vertices of the tree under construction to the corresponding integers. The transition tree of ${\cal A}_{\overline N}$ is constructed inductively as follows:
\begin{description}
\item[(A1)] The root of the tree is $\tau_\varepsilon=\tau$, and we put $T_0=\{\tau_\varepsilon\}$ and
$s(\tau_\varepsilon)=1$,
and we compute the value $\tau_{\overline N}(\tau_\varepsilon)=\sigma\circ\tau_{\varepsilon} $.
\item[(A2)] After the $i$th step let a tree $T_i$ have been constructed,
and vertices in $T_i$ have been labelled either 'closed' or 'non-closed'.
The meaning of these two terms will be made clear in the sequel.
\item[(A3)] In the next step we construct a tree $T_{i+1}$ by enriching $T_i$ in the following way: for any non-closed leaf $\tau_u$ occuring in $T_i$, where $u\in X^*$, and any $x\in X$ we add a vertex $\tau_{xu}=\delta_x\circ \tau_{u} $ and an edge from $\tau_u$ to $\tau_{xu}$ la\-belled by $x$. Simultaneously, we check whether $\tau_{xu}$ is a fuzzy set that has already been constructed. If it is true, if $\tau_{xu}$ is equal to some previously computed $\tau_v$, we mark $\tau_{xu}$ as closed and set $s(\tau_{xu})=s(\tau_{v})$. Otherwise, we compute the value $\tau_{\overline N}(\tau_{xu})=\sigma\circ\tau_{xu} $ and set $s(\tau_{xu})$ to be the next unassigned integer. The procedure terminates when all leaves are marked closed.
\item[(A4)] When the transition tree of ${\cal A}_{\overline N}$ is constructed, we erase
all closure marks and glue leaves to interior vertices with the same pointer value. The diagram that results is the transition graph of ${\cal A}_{\overline N}$.
\end{description}
\end{algorithm}

Suppose that the subsemiring ${\cal L}^*(\delta,\sigma,\tau )$ of the semiring ${\cal L}^*=(L,\lor,\otimes,0,1)$ generated by all membership values taken by $\delta $, $\sigma $ and $\tau $ is finite and has $k$ ele\-ments. Then the reverse Nerode automaton ${\cal A}_{\overline N}$ has at most $k^n$ states, and according to the analysis of computation time of algorithms provided in \cite{JMIC.14}, the computation time of Algo\-rithm \ref{alg:A.reversNerode} is $O(mnk^{2n})$. Using the same arguments we conclude that the Nerode automaton of $\cal A$ also has at most $k^n$ states.

Now we provide the following algorithm.

\begin{algorithm}[{\emph{Construction of the automaton ${\cal A}_d$}}]\label{alg:A.d}\rm The input of this algorithm is a fuzzy finite autom\-aton ${\cal A}=(A,\delta,\sigma,\tau )$ with $n$ states, over a finite alphabet $X$ with $m$ letters and a complete
residuated lattice $\cal L$, and the output is the crisp-deterministic automaton
${\cal A}_{d}=(A_{d},\delta_{d},d_{\varepsilon},\tau_{d})$.

The procedure is to construct the {\it transition tree\/} of ${\cal A}_{d}$ directly from $\cal A$, and during this procedure we use pointers $s(\cdot)$ which points vertices of the tree under construction to the corresponding integers. The transition tree of ${\cal A}_d$ is constructed inductively as follows:
\begin{description}
\item[(B1)] First we compute all members of the family $\{\tau_{w}\}_{w\in X^*}$, using steps (A1)--(A3) of Algo\-rithm \ref{alg:A.reversNerode}.
\item[(B2)] The root of the tree is $d_{\varepsilon}$, computed using formula (\ref{eq:de.varepsilon}), and we put $T_{0}=\{ d_{\varepsilon} \}$ and $s(d_\varepsilon)=1$, and we compute the value $\tau_{\overline N}(d_\varepsilon)=d_{\varepsilon}\circ\tau  $.
\item[(B3)] After the $i$th step let a tree $T_i$ have been constructed, and vertices in $T_i$ have been labelled either 'closed' or 'non-closed'. The meaning of these two terms will be made clear in the sequel.
\item[(B4)] In the next step we construct a tree $T_{i+1}$ by enriching $T_i$ in the follow\-ing way: for any non-closed leaf $d_u$ occuring in $T_i$, where $u\in X^*$, and each $x\in X$ we add a vertex $d_{ux}$ computed using formula (\ref{eq:de.ux}), and an edge from $d_u$ to $d_{ux}$ la\-belled by $x$. Simultaneously, we check whether $d_{ux}$ is a fuzzy set that has already been constructed. If it is
true, if $d_{ux}$ is equal to some previously computed $d_{v}$, we mark $d_{ux}$ as closed and set $s(d_{ux})=s(d_{v})$. Otherwise, we compute the value $\tau_d(d_{u})=d_{u}\circ\tau$
and set $s(d_{ux})$ to be the next unassigned integer. The procedure terminates when all leaves are marked closed.
\item[(B5)] When the transition tree of ${\cal A}_d$ is constructed, we erase all closure marks and glue leaves to interior vertices with the same pointer value. The diagram that results is the transition graph of ${\cal A}_d$.
\end{description}
\end{algorithm}

Assume again that the subsemiring ${\cal L}^*(\delta,\sigma,\tau )$ is finite and has $k$ ele\-ments. As we have already said, the collection $\{\tau_{w}\}_{w\in X^*}$ computed in step (B1) has at most $k^n$ different members, and the computational time of this step is $O(mnk^{2n})$. Subsequent steps produce the transition tree of a minimal crisp-deterministic fuzzy autom\-aton equivalent to $\cal A$, which can not
be larger than the Nerode automaton ${\cal A}_{N}$ of $\cal A$. As the Nerode
automaton ${\cal A}_N$ has at most $k^n$ states, the resulting transition tree for ${\cal A}_d$ has at most $k^n$ internal vertices,
and the total number of vertices does not exceed $mk^n+1$. In contrast to other determinization algorithms provided in \cite{ICB.08,JC.14,JIC.11,JMIC.14}, where the most time-demanding
part is the check whether the just compu\-ted fuzzy set is a copy of
some previously computed fuzzy set, here most of the time is spent on compu\-ting the fuzzy sets $d_u$, $u\in X^*$. Namely, we can write (\ref{eq:de.varepsilon})
as
\begin{equation}
d_{\varepsilon}(a)=\bigwedge_{\mu\in A_{\overline N}}\mu(a)\to \sigma\circ\mu,
\end{equation}
for all  $a\in A$, and (\ref{eq:de.ux}) as
\begin{equation}
d_{ux}(a)=\bigwedge_{\mu\in A_{\overline N}}\mu(a)\to d_u\circ\mu_x,
\end{equation}
for all $a\in A$, $u\in X^*$ and $x\in X$, where $\mu_x$ denotes the $x$-child of $\mu $ in the transition tree of
the reverse Nerode automaton ${\cal A}_{\overline N}$ of $\cal A$ (if $\mu=\tau_v$, for some $v\in
X^*$, then $\mu_x=\tau_{xv}$). The computation of any single $d_u\circ\mu_x$
(and, similarly, of $\sigma\circ \mu$) takes $O(n(c_\otimes +c_\lor))$
time, and therefore, the computation of the whole collection of values $\{d_u\circ\mu_x\}_{\mu\in
A_{\overline N}}$ takes $O(nk^n(c_\otimes +c_\lor))$ time, since $|A_{\overline
N}|\leqslant k^n$. When all  values $d_u\circ\mu_x$ are stored, then the
computation of $d_{ux}(a)$ requires time $O(k^n(c_\to+c_\land))$, and the
computation
of  $d_{ux}$ requires time $O(nk^n(c_\to+c_\land))$. Hence, the compu\-tation of any single vertice of the transition tree of  the automaton ${\cal A}_d$ requires time $O(nk^n(c_\otimes +c_\lor +c_\to+c_\land))$, and since there
are at most $mk^n+1$ vertices to be computed, the computation of all verices
requires $O(mnk^{2n}(c_\otimes +c_\lor +c_\to+c_\land))$ time. As for all algorithms considered in \cite{JMIC.14}, the check whether the just computed fuzzy set is a copy of some previously computed fuzzy set is performed in
time $O(mnk^{2n})$. Therefore, the whole algorithm runs in time $O(mnk^{2n}(c_\otimes +c_\lor +c_\to+c_\land))$, and if the basic operations in $\cal L$ can be performed in con\-stant time, then this algorithm has the  computational time  $O(mnk^{2n})$, the same as
all other determinization algorithms provided in \cite{ICB.08,JIC.11,JMIC.14}.

The canonization procedure given in Algorithm \ref{alg:A.d}
has the same initial stage as the Brzozowski type canonization~procedure provided in \cite{JC.14},   the computation of the family $\{\tau_{w}\}_{w\in X^*}$, i.e., the states~of~the reverse Nerode
automaton ${\cal A}_{\overline N}$.~As we have already said, the computation
time of this stage is  $O(mnk^{2n})$. In the subsequent phases these two procedures differ considerably. While in its latter stages Algorithm~\ref{alg:A.d}
  works with vectors of size $n$, where
$n$ is the number of states of $\cal A$,  the Brzozowski type procedure in its latter stages works with vectors of size $r$ and square
matrices of size $r\times r$, where~$r=|{\cal A}_{\overline N}|\leqslant
k^n$.~In~its second round the Brzozowski type procedure
produces the  reverse Nerode
automaton of ${\cal A}_{\overline N}$, which is a minimal~crisp-deter\-mi\-nistic fuzzy automaton equivalent to $\cal A$
and it is not greater than the Nerode automaton  ${\cal A}_{N}$. Thus,
this automaton does not have more than $k^n$ states, i.e., the resulting transition~tree has not more than~$k^n$~internal~vertices, and the  total number of vertices is not greater than $mk^n+1$.~The  computation of any single vertex of this transition tree requires time $O(k^{2n}(c_\otimes+c_\lor))$, so the time required to compute all vertices is $O(mk^{3n}(c_\otimes+c_\lor))$.
Since the tree has at most $mk^n$ edges, the computation time of their forming
is $O(mk^n)$.

When for any newly-constructed fuzzy set we check whether it is a copy of some previously computed fuzzy set, the total number of performed checks
is $\frac 12 k^n(k^n+1)+(m-1)k^{2n}$, and the computation time for  any single check has  $O(k^{n})$,
so the computation time for all performed checks is $O(mk^{3n})$.~Hence, the computation time of the whole algorithm is $O(mk^{3n}(c_\otimes+c_\lor))$,
or  $O(mk^{3n})$,~if~the~operations $\otimes $ and $\lor $ can be performed in constant time.~This means that the procedure proposed in Algorithm \ref{alg:A.d}
has somewhat better computation time than the Brzozowski type procedure.

Now we give a computational example that demonstrates the application of Algorithm \ref{alg:A.d}.

\begin{figure}
\begin{center}
\psset{unit=1cm}
\newpsobject{showgrid}{psgrid}{subgriddiv=1,griddots=10,gridlabels=6pt}
\begin{pspicture}(-6,-3.7)(12,4.5)
\rput(-6,4.5){\textsf{a)}}
\rput(-1.6,4.5){\textsf{b)}}
\rput(7.7,4.5){\textsf{c)}}
\rput(-4.5,-0.5){\textsf{d)}}
\rput(6,-0.5){\textsf{e)}}
\pnode(-5,2){PAP1}
\SpecialCoor
\rput(PAP1){\cnode{2.5mm}{PAP1A1}}
\rput(PAP1A1){\scriptsize$a_1$}
\rput([angle=-30,nodesep=25mm,offset=0pt]PAP1A1){\cnode{2.5mm}{PAP1A2}}
\rput(PAP1A2){\scriptsize$a_2$}
\rput([angle=30,nodesep=25mm,offset=0pt]PAP1A1){\cnode{2.5mm}{PAP1A3}}
\rput(PAP1A3){\scriptsize$a_3$}
\rput([angle=180,nodesep=5mm,offset=0pt]PAP1A1){\pnode{PAP1I}}
\ncline{->}{PAP1I}{PAP1A1}\aput[1pt](.5){\scriptsize $1$}
\rput([angle=0,nodesep=5mm,offset=0pt]PAP1A2){\pnode{PAP2I}}
\ncline{->}{PAP1A2}{PAP2I}\aput[1pt](.5){\scriptsize $1$}
\psset{nrot=:U}
\ncline{->}{PAP1A1}{PAP1A2}\bput[2pt](.5){\scriptsize $x/0.5, y/1$}
\ncline{->}{PAP1A1}{PAP1A3}\aput[1pt](.5){\scriptsize $x/1, y/0.3$}
\ncline{->}{PAP1A3}{PAP1A2}\aput[1pt](.5){\scriptsize $x/1, y/0.3$}
\psset{nrot=:D}
\nccurve[angleA=-45,angleB=-135,ncurv=4]{->}{PAP1A2}{PAP1A2}\aput[1.5pt](.50){\scriptsize $x/1, y/1$}
\nccurve[angleA=45,angleB=135,ncurv=4]{->}{PAP1A3}{PAP1A3}\bput[0.pt](.50){\scriptsize $x/0.5, y/1$}
\psset{nrot=0}
\pnode(3.3,0.5){PC}
\SpecialCoor
\rput(PC){\cnode{3mm}{PSE}}
\rput(PSE){\scriptsize$\tau_\varepsilon$}
\rput([angle=160,nodesep=22mm,offset=0pt]PC){\cnode{3mm}{PSX}}
\rput(PSX){\scriptsize$\tau_x$}
\rput([angle=20,nodesep=22mm,offset=0pt]PC){\cnode{3mm}{PSY}}
\rput(PSY){\scriptsize$\tau_y$}
\rput([angle=135,nodesep=15mm,offset=0pt]PSX){\cnode{3mm}{PSX2}}
\rput(PSX2){\scriptsize$\tau_{\!x^2}$}
\rput([angle=45,nodesep=15mm,offset=0pt]PSX){\cnode{3mm}{PSXY}}
\rput(PSXY){\scriptsize$\tau_{yx}$}
\nput[labelsep=1pt]{90}{PSXY}{$\blacksquare$}
\rput([angle=135,nodesep=15mm,offset=0pt]PSY){\cnode{3mm}{PSYX}}
\rput(PSYX){\scriptsize$\tau_{xy}$}
\nput[labelsep=1pt]{90}{PSYX}{$\blacksquare$}
\rput([angle=45,nodesep=15mm,offset=0pt]PSY){\cnode{3mm}{PSY2}}
\rput(PSY2){\scriptsize$\tau_{y^2}$}
\nput[labelsep=1pt]{90}{PSY2}{$\blacksquare$}

\rput([angle=120,nodesep=12mm,offset=0pt]PSX2){\cnode{3mm}{PSX3}}
\rput(PSX3){\scriptsize$\tau_{x^3}$}
\nput[labelsep=1pt]{90}{PSX3}{$\blacksquare$}
\rput([angle=60,nodesep=12mm,offset=0pt]PSX2){\cnode{3mm}{PSX2Y}}
\rput(PSX2Y){\scriptsize$\tau_{yx^2}$}
\nput[labelsep=1pt]{90}{PSX2Y}{$\blacksquare$}
\NormalCoor
\ncline{->}{PSE}{PSX}\aput[1pt](.6){\scriptsize $x$}
\ncline{->}{PSE}{PSY}\bput[1pt](.6){\scriptsize $y$}
\ncline{->}{PSX}{PSX2}\aput[1pt](.6){\scriptsize $x$}
\ncline{->}{PSX}{PSXY}\bput[1pt](.6){\scriptsize $y$}
\ncline{->}{PSY}{PSYX}\aput[1pt](.6){\scriptsize $x$}
\ncline{->}{PSY}{PSY2}\bput[1pt](.6){\scriptsize $y$}
\ncline{->}{PSX2}{PSX3}\aput[1pt](.5){\scriptsize $x$}
\ncline{->}{PSX2}{PSX2Y}\bput[1pt](.5){\scriptsize $y$}
\ncarc[linestyle=dashed, dash=0.7pt 2pt]{->}{PSYX}{PSX}
\ncarc[linestyle=dashed, dash=0.7pt 2pt,arcangle=-15]{->}{PSY2}{PSY}
\ncarc[linestyle=dashed, dash=0.7pt 2pt,arcangle=-14]{->}{PSXY}{PSX2}
\ncarc[linestyle=dashed, dash=0.7pt 2pt,arcangle=20]{->}{PSX3}{PSX2}
\ncarc[linestyle=dashed, dash=0.7pt 2pt,arcangle=20]{<-}{PSX2}{PSX2Y}
\pnode(9.7,0.5){PC2}
\SpecialCoor
\rput(PC2){\cnode{2.5mm}{PGSE}}
\rput(PGSE){\scriptsize$\tau_\varepsilon$}
\rput([angle=180,nodesep=5mm,offset=0pt]PGSE){\pnode{PGI}}
\ncline{->}{PGI}{PGSE}
\rput([angle=135,nodesep=15mm,offset=0pt]PGSE){\cnode{2.5mm}{PGSX}}
\rput(PGSX){\scriptsize$\tau_x$}
\rput([angle=90,nodesep=5mm,offset=0pt]PGSX){\pnode{PGSXO}}
\ncline{->}{PGSX}{PGSXO}\aput[1.5pt](.5){\scriptsize $0.5$}
\rput([angle=45,nodesep=15mm,offset=0pt]PGSE){\cnode{2.5mm}{PGSY}}
\rput(PGSY){\scriptsize$\tau_y$}
\rput([angle=90,nodesep=5mm,offset=0pt]PGSY){\pnode{PGSYO}}
\ncline{->}{PGSY}{PGSYO}\bput[1pt](.5){\scriptsize $1$}
\rput([angle=45,nodesep=15mm,offset=0pt]PGSX){\cnode{2.5mm}{PGSX2}}
\rput(PGSX2){\scriptsize$\tau_{\!x^2}$}
\rput([angle=0,nodesep=5mm,offset=0pt]PGSX2){\pnode{PGSX2O}}
\ncline{->}{PGSX2}{PGSX2O}\aput[1pt](.5){\scriptsize $1$}
\ncline{->}{PGSE}{PGSX}\aput[1pt](.5){\scriptsize $x$}
\ncline{<-}{PGSY}{PGSE}\aput[1pt](.5){\scriptsize $y$}
\psset{nrot=:U}
\ncline{->}{PGSX}{PGSX2}\aput[1pt](.5){\scriptsize $x,y$}
\psset{nrot=0}
\ncline{<-}{PGSX}{PGSY}\aput[1pt](.5){\scriptsize $x$}
\nccurve[angleA=135,angleB=45,ncurv=4]{->}{PGSX2}{PGSX2}\aput[0.5pt](.50){\scriptsize $x,y$}
\nccurve[angleA=45,angleB=-45,ncurv=4]{->}{PGSY}{PGSY}\aput[1pt](.50){\scriptsize $y$}
\pnode(1,-3.5){PC4}
\SpecialCoor
\rput(PC4){\cnode{3mm}{PSSE}}
\rput(PSSE){\scriptsize$d_\varepsilon$}
\rput([angle=160,nodesep=22mm,offset=0pt]PC4){\cnode{3mm}{PSSX}}
\rput(PSSX){\scriptsize$d_x$}
\rput([angle=20,nodesep=22mm,offset=0pt]PC4){\cnode{3mm}{PSSY}}
\rput(PSSY){\scriptsize$d_y$}
\rput([angle=135,nodesep=15mm,offset=0pt]PSSX){\cnode{3mm}{PSSX2}}
\rput(PSSX2){\scriptsize$d_{x^2}$}
\rput([angle=45,nodesep=15mm,offset=0pt]PSSX){\cnode{3mm}{PSSXY}}
\rput(PSSXY){\scriptsize$d_{xy}$}
\nput[labelsep=1pt]{90}{PSSXY}{$\blacksquare$}
\rput([angle=135,nodesep=15mm,offset=0pt]PSSY){\cnode{3mm}{PSSYX}}
\rput(PSSYX){\scriptsize$d_{yx}$}
\nput[labelsep=1pt]{90}{PSSYX}{$\blacksquare$}
\rput([angle=45,nodesep=15mm,offset=0pt]PSSY){\cnode{3mm}{PSSY2}}
\rput(PSSY2){\scriptsize$d_{y^2}$}
\nput[labelsep=1pt]{90}{PSSY2}{$\blacksquare$}
\nput[labelsep=1pt]{90}{PSSX2}{$\blacksquare$}
\NormalCoor
\ncline{->}{PSSE}{PSSX}\aput[1pt](.6){\scriptsize $x$}
\ncline{->}{PSSE}{PSSY}\bput[1pt](.6){\scriptsize $y$}
\ncline{->}{PSSX}{PSSX2}\aput[1pt](.6){\scriptsize $x$}
\ncline{->}{PSSX}{PSSXY}\bput[1pt](.6){\scriptsize $y$}
\ncline{->}{PSSY}{PSSYX}\aput[1pt](.6){\scriptsize $x$}
\ncline{->}{PSSY}{PSSY2}\bput[1pt](.6){\scriptsize $y$}
\ncarc[linestyle=dashed, dash=0.7pt 2pt,arcangle=20]{->}{PSSYX}{PSSY}
\ncarc[linestyle=dashed, dash=0.7pt 2pt,arcangle=-20]{->}{PSSY2}{PSSY}
\ncarc[linestyle=dashed, dash=0.7pt 2pt,arcangle=-20]{->}{PSSXY}{PSSX}
\ncarc[linestyle=dashed, dash=0.7pt 2pt,arcangle=-16]{->}{PSSX2}{PSSY}
\ncarc[linestyle=dashed, dash=0.7pt 2pt,arcangle=-20]{->}{PSSX2Y}{PSSX2}
\pnode(8.5,-3.2){PC5}
\SpecialCoor
\rput(PC5){\cnode{2.5mm}{PGGSE}}
\rput(PGGSE){\scriptsize$d_\varepsilon$}
\rput([angle=180,nodesep=5mm,offset=0pt]PGGSE){\pnode{PGGI}}
\ncline{->}{PGGI}{PGGSE}
\rput([angle=135,nodesep=15mm,offset=0pt]PGGSE){\cnode{2.5mm}{PGGSX}}
\rput(PGGSX){\scriptsize$d_x$}
\rput([angle=90,nodesep=5mm,offset=0pt]PGGSX){\pnode{PGGSXO}}
\ncline{->}{PGGSX}{PGGSXO}\bput[1.5pt](.5){\scriptsize $0.5$}
\rput([angle=45,nodesep=15mm,offset=0pt]PGGSE){\cnode{2.5mm}{PGGSY}}
\rput(PGGSY){\scriptsize$d_y$}
\rput([angle=90,nodesep=5mm,offset=0pt]PGGSY){\pnode{PGGSYO}}
\ncline{->}{PGGSY}{PGGSYO}\aput[1.5pt](.5){\scriptsize $1$}
\ncline{->}{PGGSE}{PGGSX}\aput[1pt](.5){\scriptsize $x$}
\ncline{<-}{PGGSY}{PGGSE}\aput[1pt](.5){\scriptsize $y$}
\ncline{->}{PGGSX}{PGGSY}\aput[1pt](.5){\scriptsize $x$}
\nccurve[angleA=135,angleB=225,ncurv=4]{->}{PGGSX}{PGGSX}\bput[1pt](.50){\scriptsize $y$}
\psset{nrot=:D}
\nccurve[angleA=45,angleB=-45,ncurv=4]{->}{PGGSY}{PGGSY}\aput[1pt](.50){\scriptsize $x,y$}
\psset{nrot=0}
\NormalCoor
\end{pspicture}\\
\renewcommand{\figurename}{\scriptsize Fig.}
\caption{\scriptsize The transition graph of the fuzzy automaton ${\cal A}$ from Example \ref{ex:ex1} (a)), the transition tree (b)) and the transition graph (c)) of the reverse Nerode automaton ${\cal A}_{\overline N}$, and the transition tree (d)) and the transition graph (e)) of the automaton ${\cal A}_d$}\label{fig:ffaA}
\end{center}
\end{figure}
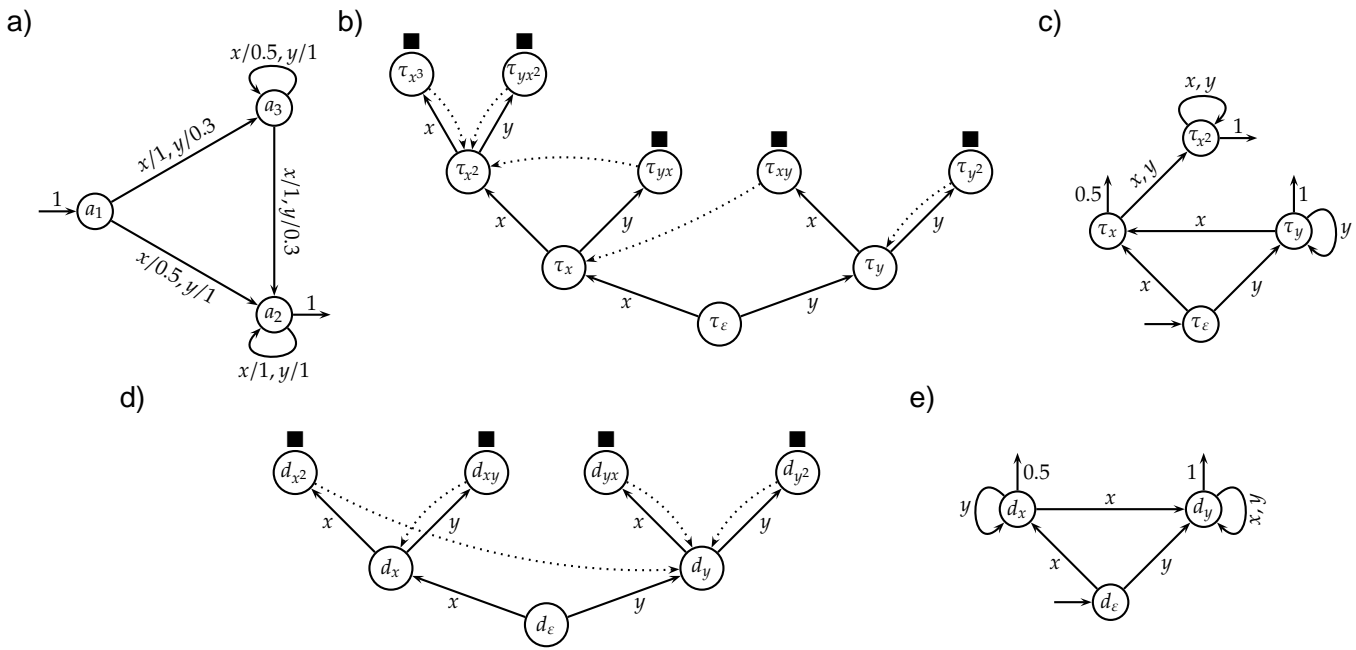

\begin{example}\label{ex:ex.prod}\rm
Let ${\cal A}=(A,\delta,\sigma,\tau)$ be a fuzzy finite automaton over the two-element alphabet $X=\{x,y\}$ and the Goguen (product) structure, given by the transition graph shown in Fig.~\ref{ex:ex.prod} a). Equivalently, $\cal A$  is specified by the following matrices and vectors:
\[
\begin{aligned}
&\delta_x=\begin{bmatrix}
0 & 0.5 & 1 \\
0 & 1 & 0 \\
0 & 1 & 0.5 \\

\end{bmatrix},\ \ \
\delta_y=\begin{bmatrix}
0 &  1 & 0.3 \\
0 & 1 & 0 \\
0 & 0.3  & 1
\end{bmatrix},\ \ \
&\sigma=\begin{bmatrix}
1 &
0 &
0
\end{bmatrix},\ \ \
\tau=\begin{bmatrix}
0\\
1  \\
0
\end{bmatrix}.
\end{aligned}
\]
Using step (B1) of Algorithm \ref{alg:A.d} (i.e., Algorithm \ref{alg:A.reversNerode}) we obtain that the family $\{\tau_{w}\}_{w\in X^*}$ consists of fuzzy sets
represented by the following vectors:
\[
\tau_\varepsilon=\begin{bmatrix}0 \\ 1 \\ 0 \end{bmatrix}, \ \
\tau_x=\begin{bmatrix} 0.5 \\ 1 \\ 1 \end{bmatrix}, \ \
\tau_y=\begin{bmatrix} 1 \\ 1 \\ 0.3 \end{bmatrix}, \ \
\tau_{x^2}=\begin{bmatrix} 1 \\ 1 \\ 1 \end{bmatrix},
\]
since $\tau_{xy}=\tau_x,\ \tau_{yx}=\tau_{x^2}=\tau_{x^3}=\tau_{yx^2},\ \tau_{y^2}=\tau_y$.~The transition tree and transition graph of the reverse Nerode automaton ${\cal A}_{\overline N}$ are given by Fig.~\ref{ex:ex.prod} b) and c).~In (B2) we first compute products $\sigma\circ \tau_\varepsilon=0$, $\sigma\circ \tau_x=0.5$, $\sigma\circ \tau_y=1$ and $\sigma\circ \tau_{x^2}=1$, and then, for any $a\in A$, we obtain that
\[
\begin{aligned}
d_\varepsilon (a)&=(\tau_\varepsilon (a)\to \sigma\circ \tau_\varepsilon)\land (\tau_x (a)\to \sigma\circ \tau_x)\land (\tau_y (a)\to \sigma\circ \tau_y)\land (\tau_{x^2}(a)\to \sigma\circ \tau_{x^2})= \\
&= (\tau_\varepsilon (a)\to 0)\land (\tau_x (a)\to 0.5)\land (\tau_y (a)\to 1)\land (\tau_{x^2}(a)\to 1)=(\tau_\varepsilon (a)\to 0)\land (\tau_x (a)\to 0.5),
\end{aligned}
\]
whence it follows that
\[
d_\varepsilon=\begin{bmatrix} 1 \\ 0 \\ 0.5 \end{bmatrix}, \ \ d_\varepsilon\circ \tau_\varepsilon=0, \ \ d_\varepsilon\circ \tau_x=0.5, \ \ d_\varepsilon\circ \tau_y=1, \ \ d_\varepsilon\circ \tau_{x^2}=1 .
\]
In (B4), for each $a\in A$ we have that
\[
\begin{aligned}
d_x(a)&=(\tau_\varepsilon (a)\to d_\varepsilon\circ \tau_x)\land (\tau_x (a)\to d_\varepsilon\circ \tau_{x^2})\land (\tau_y (a)\to d_\varepsilon\circ \tau_{x})\land (\tau_{x^2}(a)\to d_\varepsilon\circ \tau_{x^2})= \\
&= (\tau_\varepsilon (a)\to 0.5)\land (\tau_x (a)\to 1)\land (\tau_y (a)\to 0.5)\land (\tau_{x^2}(a)\to 1)= (\tau_\varepsilon (a)\to 0.5) \land (\tau_y (a)\to 0.5), \\
d_y(a)&=(\tau_\varepsilon (a)\to d_\varepsilon\circ \tau_y)\land (\tau_x (a)\to d_\varepsilon\circ \tau_{x^2})\land (\tau_y (a)\to d_\varepsilon\circ \tau_{y})\land (\tau_{x^2}(a)\to d_\varepsilon\circ \tau_{x^2})= \\
&= (\tau_\varepsilon (a)\to 1)\land (\tau_x (a)\to 1)\land (\tau_y (a)\to 1)\land (\tau_{x^2}(a)\to 1)=1,
\end{aligned}
\]
and therefore,
\[
d_x=\begin{bmatrix} 0.5 \\ 0.5 \\ 1 \end{bmatrix}, \ \ d_y=\begin{bmatrix} 1 \\ 1 \\ 1 \end{bmatrix}, \ \
\begin{matrix}
d_x\circ \tau_\varepsilon=0.5, \ \ d_x\circ \tau_x=1, \ \ d_x\circ \tau_y=0.5, \ \ d_x\circ \tau_{x^2}=1 , \\
d_y\circ \tau_\varepsilon=1, \ \ d_y\circ \tau_x=1, \ \ d_y\circ \tau_y=1, \ \ d_y\circ \tau_{x^2}=1 .
\end{matrix}
\]
Next, for each $a\in A$ we have
\[
\begin{aligned}
d_{x^2}(a)&=(\tau_\varepsilon (a)\to d_x\circ \tau_x)\land (\tau_x (a)\to d_x\circ \tau_{x^2})\land (\tau_y (a)\to d_x\circ \tau_{x^2})\land (\tau_{x^2}(a)\to d_x\circ \tau_{x^2})= \\
&= (\tau_\varepsilon (a)\to 1)\land (\tau_x (a)\to 1)\land (\tau_y (a)\to 1)\land (\tau_{x^2
}(a)\to 1)=1=d_{y}(a), \\
d_{xy}(a)&=(\tau_\varepsilon (a)\to d_x\circ \tau_y)\land (\tau_x (a)\to d_x\circ \tau_{x^2})\land (\tau_y (a)\to d_x\circ \tau_{y})\land (\tau_{x^2}(a)\to d_x\circ \tau_{x^2})= \\
&= (\tau_\varepsilon (a)\to 0.5)\land (\tau_x (a)\to 1)\land (\tau_y (a)\to 0.5)\land (\tau_{x^2}(a)\to 1)=  \\
&= (\tau_\varepsilon (a)\to 0.5)\land (\tau_y (a)\to 0.5)=d_{x}(a),  \\
d_{yx}(a)&=(\tau_\varepsilon (a)\to d_y\circ \tau_x)\land (\tau_x (a)\to d_y\circ \tau_{x^2})\land (\tau_y (a)\to d_y\circ \tau_{x})\land (\tau_{x^2}(a)\to d_y\circ \tau_{x^2})= \\
&= (\tau_\varepsilon (a)\to 1)\land (\tau_x (a)\to 1)\land (\tau_y (a)\to 1)\land (\tau_{yx}(a)\to 1)= 1=d_{y}(a), \\
d_{y^2}(a)&=(\tau_\varepsilon (a)\to d_y\circ \tau_y)\land (\tau_x (a)\to d_y\circ \tau_{x^2})\land (\tau_y (a)\to d_y\circ \tau_{y})\land (\tau_{x^2}(a)\to d_y\circ \tau_{x^2})= 1=d_{y}(a)
\end{aligned}
\]
so
$d_{x^2}=d_{yx}=d_{y^2}=d_y$ and $d_{xy}=d_x$, and our algorithm terminates.~Its results are the transition tree of the automaton ${\cal A}_d$, represented by Fig.~\ref{ex:ex.prod} d), and the transition graph of ${\cal A}_d$, represented by Fig.~\ref{ex:ex.prod} e).

Therefore, the minimal crisp-deterministic fuzzy automaton equivalent to the fuzzy finite automaton $\cal A$ from this example has 3 states. It is easy to check that algorithms developed in \cite{ICB.08,JIC.11}, applied to the same fuzzy finite automaton $\cal A$, yield infinite crisp-deterministic fuzzy automata.

Note that in this case our method gave a finite crisp-deterministic automaton regardless of the fact~that the subsemiring ${\cal L}^*(\delta,\sigma,\tau )$ is infinite.
\end{example}

We also give another example.
\begin{example}\label{ex:ex1}
Let ${\cal A}=(A,\delta,\sigma,\tau)$ be a Boolean automaton over the two-element alphabet $X=\{x,y\}$~given by the transition graph shown in Fig. \ref{fig:ffaA} a). 
\begin{figure}
\begin{center}
\psset{unit=1cm}
\newpsobject{showgrid}{psgrid}{subgriddiv=1,griddots=10,gridlabels=6pt}
\begin{pspicture}(-4,0)(7,4)
\rput(-4,4){\textsf{a)}}
\rput(2.5,4){\textsf{b)}}
\pnode(-3,2){AP1}
\SpecialCoor
\rput(AP1){\cnode[doubleline=true]{3mm}{AP1A1}}
\rput(AP1A1){\scriptsize$a_1$}
\rput([angle=-30,nodesep=25mm,offset=0pt]AP1A1){\cnode{2.7mm}{AP1A2}}
\rput(AP1A2){\scriptsize$a_2$}
\rput([angle=30,nodesep=25mm,offset=0pt]AP1A1){\cnode[doubleline=true]{3mm}{AP1A3}}
\rput(AP1A3){\scriptsize$a_3$}
\rput([angle=180,nodesep=5mm,offset=0pt]AP1A1){\pnode{AP1I}}
\ncline{->}{AP1I}{AP1A1}
\ncarc[arcangle=16]{<-}{AP1A1}{AP1A2}\aput[1pt](.5){\scriptsize $x,y$}
\ncarc[arcangle=16]{<-}{AP1A2}{AP1A1}\aput[1pt](.5){\scriptsize $x$}
\ncarc[arcangle=16]{<-}{AP1A1}{AP1A3}\aput[1pt](.5){\scriptsize $x$}
\ncarc[arcangle=16]{<-}{AP1A3}{AP1A1}\aput[1pt](.5){\scriptsize $y$}
\ncarc[arcangle=16]{<-}{AP1A2}{AP1A3}\aput[1pt](.5){\scriptsize $y$}
\ncarc[arcangle=16]{<-}{AP1A3}{AP1A2}\aput[1pt](.5){\scriptsize $x$}
%
\nccurve[angleA=45,angleB=-45,ncurv=4]{->}{AP1A2}{AP1A2}\aput[0.5pt](.50){\scriptsize $y$}
\pnode(5,0.5){C5}
\SpecialCoor
\rput(C5){\cnode[doubleline=true]{3.5mm}{GGSE}}
\rput(GGSE){\scriptsize$d_0$}
\rput([angle=180,nodesep=5mm,offset=0pt]GGSE){\pnode{GGI}}
\ncline{->}{GGI}{GGSE}
\rput([angle=135,nodesep=15mm,offset=0pt]GGSE){\cnode{3mm}{GGSX}}
\rput(GGSX){\scriptsize$d_x$}
\rput([angle=45,nodesep=15mm,offset=0pt]GGSE){\cnode[doubleline=true]{3.5mm}{GGSY}}
\rput(GGSY){\scriptsize$d_y$}
\rput([angle=45,nodesep=15mm,offset=0pt]GGSX){\cnode[doubleline=true]{3.5mm}{GGSX2}}
\rput(GGSX2){\scriptsize$d_{x^2}$}
\ncline{->}{GGSE}{GGSX}\aput[1pt](.5){\scriptsize $x$}
\ncarc[arcangle=12]{<-}{GGSY}{GGSE}\aput[1pt](.5){\scriptsize $y$}
\ncline{->}{GGSX}{GGSX2}\aput[1pt](.5){\scriptsize $x,y$}
\ncline{<-}{GGSX}{GGSY}\aput[1pt](.5){\scriptsize $y$}
\ncarc[arcangle=-12]{->}{GGSY}{GGSE}\bput[1pt](.5){\scriptsize $x$}
\nccurve[angleA=135,angleB=45,ncurv=4]{->}{GGSX2}{GGSX2}\aput[0.5pt](.50){\scriptsize $x,y$}
\NormalCoor
\end{pspicture}\\
\renewcommand{\figurename}{\scriptsize Fig.}
\caption{\scriptsize The transition graph of the fuzzy automaton ${\cal A}$ from Example \ref{ex:ex1} (a)), and the transition  graph  of the automaton ${\cal A}_d$ (b)).}\label{fig:ffaA}
\end{center}
\end{figure}
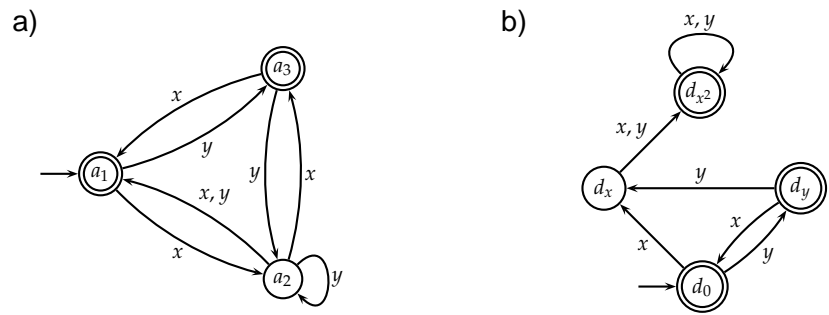
As in the previous example we construct the automaton ${\cal A}_d$, whose
transition graph is represented by Fig. \ref{fig:ffaA} b). Therefore, the
automaton ${\cal A}_d$, i.e., the minimal crisp-deterministic fuzzy automaton equivalent to  $\cal A$  has 4 states. Note that all algorithms developed in \cite{ICB.08,JIC.11,JMIC.14}, applied to the same fuzzy finite automaton $\cal A$, yield crisp-deterministic fuzzy automata with 7 and 6 states (cf. Example 4.9 in \cite{JMIC.14}).
\end{example}

\end{document}